\documentclass[aps,prl,showpacs,nobibnotes,twocolumn,preprintnumbers,
nobalancelastpage]{revtex4}        
\usepackage{graphics}
\usepackage{graphicx}

\begin{document} 
\title{Void Ellipticity Distribution as a Probe of Cosmology}
\author{Daeseong Park and Jounghun Lee}
\email{pds2001@astro.snu.ac.kr, jounghun@astro.snu.ac.kr}
\affiliation{Department of Physics and Astronomy, FPRD, Seoul National 
University, Seoul 151-747, Korea}

\begin{abstract} 
Cosmic voids refer to the large empty regions in the universe with a very low 
number density of galaxies. Voids are likely to be severely disturbed by the 
tidal effect from the surrounding dark matter. We derive a completely analytic 
model for the void ellipticity distribution from physical principles. 
We use the spatial distribution of galaxies in a void as a measure of its 
shape, tracking the trajectory of the void galaxies under the influence of 
the tidal field using the Lagrangian perturbation theory. Our model implies 
that the void ellipticity distribution depends sensitively on the cosmological 
parameters. Testing our model against the high-resolution Millennium Run 
simulation, we find excellent quantitative agreements of the analytic 
predictions with the numerical results.
\end{abstract}

\maketitle

{\it Introduction.---}One of the most fundamental goals in physical cosmology 
is to understand the large-scale structure in the universe. The organization 
of the large-scale structure begins at the scale of galaxies and appears to 
follow a hierarchical pattern up to the scale of superclusters 
\cite{col-etal01}.   

The study of the large-scale structure has been so far focused mostly on 
the bound objects like clusters and superclusters. However, more recent 
observations indicate that the universe is in fact a collection of bubble-like 
voids separated by sheets and filaments, at the dense nodes of which the 
clusters and superclusters are located \cite{hoy-vog04}. Hence, to address 
the large scale structure in the universe, it seems essential to account 
for the presence and characterization of voids.

According to the standard gravitational instability theory, voids 
originate from the local minima of the initial density field and expand 
faster than the rest of the universe. The extremely low-density of the 
observed voids ($\delta_{v} \approx -0.9$) supports this scenario 
\cite{hoy-vog04}. A common expectation based on this standard 
scenario is that voids are likely to have quite spherical shapes 
\cite{ick84,dub-etal93,van-van93,she-van04}. However, a recent systematic 
analysis of simulation data has revealed that the shapes of voids are 
in fact far from spherical symmetry \cite{sha-etal04,sha-etal06}. 

Confronted with this somewhat unexpected result, Ref.\cite{sha-etal06} 
claimed that the gravitational tidal field be responsible for the 
nonspherical shapes of voids. They explained that since the voids have 
very low-density, they must be more easily disturbed by the tidal 
effect from the surrounding matter. 

Reference \cite{lee-par06} has investigated the tidal effect on voids 
analytically using the linear tidal torque theory, which was inspired by 
\cite{sha-etal06}. It is shown by them that the tidal field indeed has a 
significant influence on voids, generating non-radial motions of matter that 
make up voids. They tested their analytic model against high-resolution 
N-body simulation and found remarkably good agreements between the analytical 
and the numerical results. 

The success of the analytic model of \cite{lee-par06} has two crucial 
implications. First, the non-radial motions of galaxies in voids  
generated by the tidal effect would cause the nonsphericity in void shapes, 
as hinted first by \cite{sha-etal06}.  Second, since the Lagrangian theory 
has turn out to work well in predicting the tidal effect on voids, 
it may be also possible to model the nonspherical shapes using the 
Lagrangian theory that depends only on the initial conditions.

One difficulty in modeling the nonspherical shapes of voids, however, 
lies in the fact that there is no well defined way to determine the shape 
of a void  since a void is an unbound system having no clear-cut boundary.
Nevertheless, given that our purpose is not to construct the full geometrical 
shape of a void but to quantify the degree of the deviation of its shape 
from spherical symmetry, a practical way that we choose here is to use the 
spatial distribution of the void galaxies as a measure of the nonspherical 
shape of a void. If a void has a spherical shape, then the spatial 
distribution of its galaxies will show a more or less isotropic pattern. 
If it has a nonspherical shape, the void galaxy spatial distribution will 
also deviate from the isotropic pattern. In this practical way, the 
nonsphericity of the shape of a given void may be quantified by the 
{\it ellipticity} of the spatial distribution of the galaxies that it 
contains. 

In this Letter we aim for constructing a complete analytic model for the 
void ellipticity using the Zel'dovich approximation (the first order 
Lagrangian perturbation theory) and investigate if the void ellipticity 
distribution can put constraints on the cosmological parameters. 

{\it The Analytic Model---}Assuming that the density field is smoothed on the 
void scale, we use the Zel'dovich approximation \cite{zel70} to describe the 
trajectory of a galaxy located in a void region as 
${\bf x}={\bf q}-D(t)\nabla\Psi({\bf q})$. Here ${\bf x}$ and ${\bf q}$ 
represent the Eulerian and the Lagrangian positions of a void galaxy, 
respectively, $D(t)$ is the linear growth factor and $\Psi({\bf q})$ is 
the perturbation potential smoothed on the Lagrangian void scale, 
$R_{L}$. Note that here a galaxy is treated as a particle moving under 
the influence of $\Psi$.

Applying the mass conservation law, 
$\rho d^{3}{\bf x} =\bar{\rho}d^{3}{\bf q}$ to the Zel'dovich approximation, 
the mass density of a void region, $\rho_{v}$, can be written as 
\begin{equation}
\rho_{v} = \frac{\bar{\rho}}{(1-\lambda_1)(1-\lambda_2)(1-\lambda_3)}, 
\label{eqn:rho}
\end{equation}
where $\lambda_{1},\lambda_{2},\lambda_{3}$ 
(with $\lambda_{1}>\lambda_{2}>\lambda_{3}$) are the three eigenvalues of 
the tidal tensor, $T_{ij}$, defined as the second derivative of $\Psi$. 
The dimensionless residual overdensity of the void region, $\delta_{v}$, 
equals the sum of the three eigenvalues: 
$\delta_{v}=\sum_{i=1}^{3}\lambda_{i}$. 

Eq.~(\ref{eqn:rho}) implies that under the influence of the tidal field, 
$T_{ij}$, the void region will experience a triaxial expansion. 
Let $p_{1},p_{2},p_{3}$ (with $p_{1}>p_{2}>p_{3}$) represent the three 
semi-axis lengths of the inertia momentum tensor of the galaxies that make 
up void , and let $\mu$ and $\nu$ be the two axial ratios defined as 
$\mu\equiv p_{2}/p_{1}$ and $\nu\equiv p_{3}/p_{1}$. The ellipticity 
of a given void region in terms of the minor-to-major axis ratio as 
$\varepsilon \equiv 1 -\nu$. 

According to Eq.~(\ref{eqn:rho}), the two axial ratios are related to the 
three eigenvalues of $(T_{ij})$ as 
\begin{equation}
\label{eqn:axis_lambda}
\mu = \left(\frac{1-\lambda_2}{1-\lambda_3}\right)^{1/2},\qquad 
\nu = \left(\frac{1-\lambda_1}{1-\lambda_3}\right)^{1/2}.
\end{equation}
The distribution of $\varepsilon$ can be found from the distribution of $\nu$, 
which in turn can be derived from the conditional distribution of the three 
eigenvalues of the tidal tensor under the constraint of 
$\delta_{v}=\sum_{i=1}^{3}\lambda_{i}$.

In Ref.\cite{dor70}, the unconditional joint probability density distribution 
of $\{\lambda_1,\lambda_2,\lambda_3\}$ was derived. On the void scale 
$R_{L}$, it is written as 
\begin{eqnarray}
p(\lambda_1,\lambda_2,\lambda_3;\sigma_{R_L})&=&
\frac{3375}{8\sqrt{5}\pi\sigma^6_{R_L}}
\exp\bigg{(}-\frac{3K_{1}^2}{\sigma^2_{R_L}}
+\frac{15K_{2}}{2\sigma^2_{R_L}}\bigg{)}\cr
&&\times(\lambda_{1}-\lambda_{2})(\lambda_{2}-\lambda_{3})
(\lambda_{1}-\lambda_{3}),
\label{eqn:dor}
\end{eqnarray}
where $K_{1} \equiv \lambda_{1}+\lambda_{2}+\lambda_{3}$, 
$K_{2} \equiv \lambda_{1}\lambda_{2} + \lambda_{2}\lambda_{3} +
\lambda_{3}\lambda_{1}$, and $\sigma_{R_L}$ is the rms density fluctuation 
related to the dimensionless linear power spectrum $\Delta^{2}(k)$ as 
\begin{equation}
\sigma^{2}_{R_L} \equiv \int_{-\infty}^{\infty}\Delta^{2}(k)
W^{2}(kR_{L})d\ln k, 
\end{equation}
where $W(kR_L)$ represents a smoothing function with the Lagrangian filtering 
radius of $R_L$. Here, we adopt the approximation formula given by 
\cite{bar-etal86} for $\Delta^{2}(k)$ and the top-hat spherical filter 
for $W(kR_{L})$. 

By Eq.~(\ref{eqn:dor}), the conditional joint probability 
density distribution of the void axial ratios is found as  
\begin{eqnarray}
p(\mu,\nu|\delta_{v};\sigma_{R_L})&=& 
A~p[\lambda_{1}(\mu,\nu),\lambda_{2}(\mu,\nu)|\delta_{v};\sigma_{R_L}]
\cr
&& \times\frac{4(\delta_{v}-3)^2\mu\nu}{(\mu^{2}+\nu^{2}+1)^{3}}.
\label{eqn:munu}
\end{eqnarray}
Here $p(\lambda_{1},\lambda_{2}|\delta_{v};\sigma_{R_L})$ is the conditional 
probability density of $\lambda_{1}$ and $\lambda_{2}$ provided that a given 
void has an average density $\delta_{v}$ on the scale of $R_{L}$.  This 
conditional probability has the following form:
\begin{eqnarray} 
&&p(\lambda_{1},\lambda_{2}|\delta_{v};\sigma_{R_L}) \cr
&&=\frac{3375\sqrt{2}}{\sqrt{10\pi}\sigma^{5}_{R_L}}
\exp\left[-\frac{5\delta^{2}_{v}}{2\sigma^{2}_{R_L}} + 
\frac{15\delta_{v}(\lambda_{1}+\lambda_{2})}{2\sigma^{2}_{R_L}}\right]\cr
&&\times\exp\left[-\frac{15(\lambda^{2}_{1}+\lambda_{1}\lambda_{2}+
\lambda^{2}_{2})}{2\sigma^{2}_{R_L}}\right] \cr
&&\times(2\lambda_{1}+\lambda_{2}-\delta_{v})(\lambda_{1}-\lambda_{2})
(\lambda_{1}+2\lambda_{2}-\delta_{v}).
\label{eqn:con}
\end{eqnarray}
where the functional forms of $\lambda_{1}(\mu,\nu)$ and 
$\lambda_{2}(\mu,\nu)$ are given as \cite{lee-etal05}
\begin{eqnarray}
\label{eqn:lamu1}
\lambda_{1}(\mu,\nu) &=& \frac{1 + (\delta_{v}- 2)\nu^{2} + 
\mu^{2}}{(\mu^{2} + \nu^{2} + 1)},\\
\label{eqn:lamu2} 
\lambda_{2}(\mu,\nu) &=& \frac{1 + (\delta_{v}- 2)\mu^{2} + \nu^{2}}
{(\mu^{2} + \nu^{2} + 1)},
\end{eqnarray}
and the normalization constant, $A$, satisfies the constraint of 
$A\int p(\mu,\nu | \delta=\delta_v ; \sigma_{R_L})d\mu d\nu = 1$

The distribution, $p(\nu)$ is now straightforwardly evaluated by integrating 
Eq.~(\ref{eqn:munu}) over 
$\mu$ as 
\begin{equation}
p(\nu; R_{L}) = \int_{\nu}^{1}
p(\mu,\nu|\delta =\delta_{v};\sigma_{R_L})d\mu.
\label{eqn:nu}
\end{equation}
Then, the distribution of the void ellipticity, $p(\varepsilon)$, is nothing 
but $p(\epsilon)=p(1-\nu)$.
\begin{figure}
\includegraphics[scale=0.45]{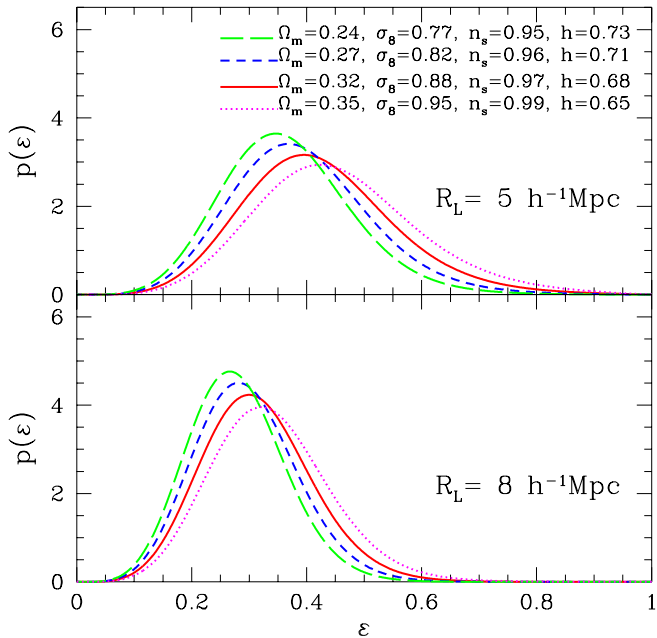}
\caption
{Probability density distributions of the void ellipticity  predicted by 
our analytic model for the four different sets of cosmological parameters 
at two different smoothing scales: $R_{L}=5$ and $8h^{-1}$Mpc in the upper 
and the lower panel, respectively. For each case, the void density contrast 
is set at $\delta_{v}=-0.9$.}
\label{fig:analytic}
\end{figure}
Fig.~\ref{fig:analytic} plots $p(\varepsilon)$ at two different Lagrangian 
scales: $R_{L}=5$ and $8h^{-1}$Mpc in the upper and the lower panel, 
respectively. Here we assume a $\Lambda$-dominated cold dark matter cosmology 
($\Omega_{\Lambda}=1-\Omega_{m}$) and consider four different sets of 
the key cosmological parameters: $\Omega_{m}$ 
(matter density parameter), $h$ (dimensionless Hubble parameter), 
$\sigma_{8}$ (amplitude), and $n_{s}$ (power spectrum slope). 

As can be seen, the ellipticity distribution depends quite sensitively on 
the cosmological parameters. Note that for the cases of low $\Omega_{m}$ 
and high $h$, voids tend to have more spherical shapes, which in fact can 
be understood as follows. The void ellipticity distribution is an outcome 
of the counterbalance between the tidal effect of the dark matter and 
the expansion of the universe: The tidal effect disturbs the void shapes 
from spherical symmetry, while the expansion of the universe resists such 
disturbance.It is also worth noting that the void ellipticity distribution 
moves toward the low ellipticity as the smoothing scale increases. It 
indicates that the larger a void is, the more spherical it is.

{\it Numerical Test.---}To examine the validity of our analytic model, 
we test its prediction against the Millennium Run simulation for which the 
cosmological parameters are specified as 
$\Omega_{m}=0.25$, $n_{s}=1$, $\sigma_{8}=0.9$ and $h=0.73$ \citep{spr-etal05}.

In our previous work \cite{lee-par06} we have already identified voids 
in the Millennium Run catalog using the Hoyle-Vogeley (HV02) void-finder 
algorithm \cite{hoy-vog02}. For the detailed description of the void-finding  
scheme and the properties of the identified voids, see \cite{lee-par06}. 
Among these voids, we select only those voids which contain 
more than $5$ galaxies. A total of $23652$ voids are found to contain more 
than $5$ galaxies.

For each selected void, we first calculate the pseudo inertia momentum 
tensor, $(I_{ij})$, defined as 
\begin{equation}
I_{ij} = \frac{1}{N_{vg}}\sum_{\alpha}x^{g}_{\alpha i}x^{g}_{\alpha j},
\label{eqn:iner}
\end{equation}
where $N_{vg}$ is the number of galaxies belonging to a given void, and 
$(x_{\alpha i}^{g})$ is the position vector of the $\alpha$-th void galaxy. 
As mentioned in \cite{sha-etal06}, the inertia momentum tensor is computed 
without weighting the position vectors with mass since we are interested 
in the geometry of voids.

Diagonalizing $(I_{ij})$, we find the three eigenvalues, $I_{1}$, $I_{2}$, 
and $I_{3}$ (with $I_{1}>I_{2}>I_{3}$), which are related to the semi 
axis-lengths of a given void as 
$I_{1}\propto (p_{1}^{2}+p_{2}^{2})$, $I_{2}\propto (p_{1}^{2}+p_{3}^{2})$, 
$I_{3}\propto (p_{2}^{2}+p_{3}^{2})$ \cite{sha-etal06}. Now, the void 
ellipticity, $\varepsilon \equiv 1-p_{3}/p_{1}$, can be written in terms of 
$I_{1},I_{2},I_{3}$ as 
\begin{equation}
\varepsilon = 1 -\nu = 1 -
\left(\frac{I_{2}+I_{3}-I_{1}}{I_{1}+I_{2}-I_{3}}\right)^{1/2}.
\label{eqn:ell}
\end{equation}
Measuring $\varepsilon$ of each selected void using Eq.~(\ref{eqn:ell}), 
we determine the void ellipticity distribution. Since the distribution 
depends on the size of void, we divide the whole sample of voids into the 
four bins according to their effective radius, $R_{E}$, which is defined as 
$R_{E}\equiv (3\Gamma_{V}/4\pi)^{1/3}$ where $\Gamma_{V}$ is the total volume 
of a void measured by the Monte-Carlo method devised by \cite{hoy-vog02}. 
Each bin is chosen to include equal number of voids. The range of $R_{E}$, 
the mean values of $R_{E}$, $\delta_{v}$, and $\varepsilon$ for the four 
bins are listed in Table ~\ref{tab:list}.
\begin{table}
\caption{The range of the effective radius ($R_{\textrm{E}}^{\textrm{min}}$ 
and $R_{\textrm{E}}^{\textrm{max}}$), the mean effective radius 
($\bar{R}_{\textrm{E}}$), the mean density ($\bar{\delta}_{\textrm{v}}$), 
the mean ellipticity ($\bar{\varepsilon})$, and the mean number of the 
galaxies ($\bar{N}_{\textrm{vg}}$) of the voids belonging to the four bins.
\label{tab:list}}
\begin{ruledtabular}
\begin{tabular}{lcccccc}
& $R_{\textrm{E}}^{\textrm{min}}$ &
$R_{\textrm{E}}^{\textrm{max}}$& $\bar{R}_{\textrm{E}}$&
$\bar{\delta}_{\textrm{v}}$
& $\bar{\varepsilon}$ & $\bar{N}_{\textrm{vg}}$ \\
\hline
$(a)$ & $6.42$  & $9.04$  & $8.37$& $-0.89$ & $0.48$ & $18$ \\
$(b)$ & $9.04$  & $10.28$ & $9.65$& $-0.89$ & $0.43$ & $28$ \\
$(c)$ & $10.28$ & $11.66$ & $10.92$ & $-0.89$ & $0.41$ & $39$ \\
$(d)$ & $11.66$ & $21.52$ & $13.00$& $-0.90$ & $0.36$ & $63$ \\
\end{tabular}
\end{ruledtabular}
\end{table}
\begin{figure} 
\includegraphics[scale=0.45]{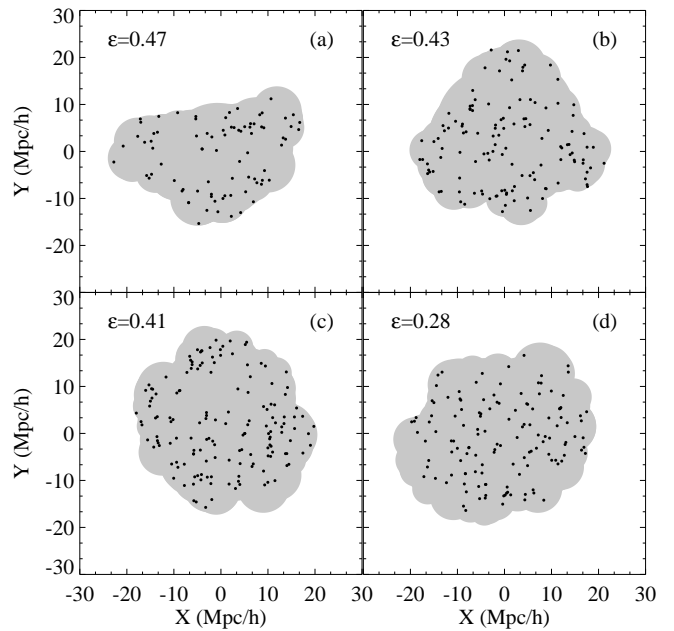}
\caption
{Illustration of the four examples of the voids identified in the Millennium 
Run simulation. Each panel corresponds to the four bins. The shaded regions 
and the dots represent the areas of the voids and the positions of the void 
galaxies in them projected onto the two dimensional plane.} 
\label{fig:exa}
\end{figure}
\begin{figure}
\includegraphics[scale=0.45]{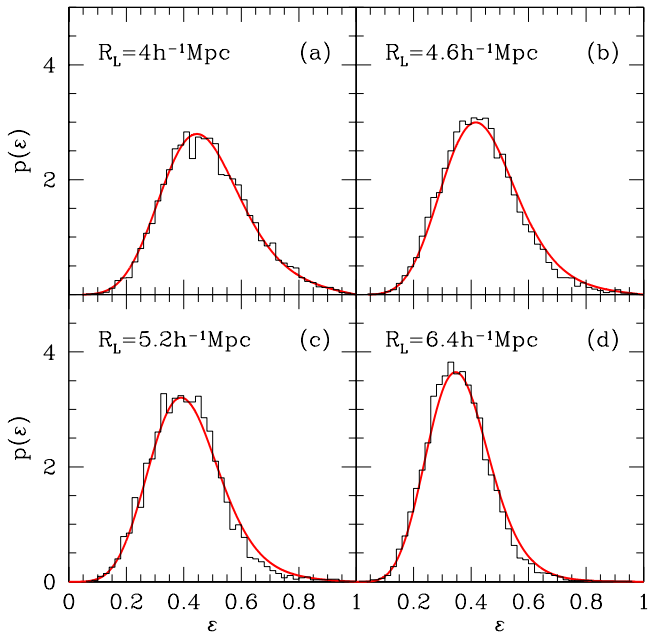}
\caption{The ellipticity distributions of the voids belonging to the 
four bins. In each panel the histogram represents the numerical data points 
from the Millennium Run simulation while the solid line corresponds to the 
analytic prediction. For the analytic results, the values of the 
Lagrangian scale radius, $R_{L}$, are evaluated by the relation of 
$R_{L}=(1+\bar{\delta}_{v})^{1/3}\bar{R}_{E}$. }
\label{fig:com}
\end{figure}
As shown, $\bar{\varepsilon}$ decreases as the mean $\bar{R}_{E}$ increases. 
That is, the larger a void is, the more spherical shape it tends to have.

Fig.~\ref{fig:exa} illustrates the projected images of the four sample voids 
selected from the four bins, respectively. In each panel the dots represent 
the positions of the void galaxies and the shaded regions correspond to 
the areas of the voids. Fig.~\ref{fig:com} plots the void ellipticity 
distributions for the four bins. In each panel, the histogram and the solid 
line represent the numerical and the analytical results, respectively. 
For the analytic distribution,  the Lagrangian scale radius, $R_{L}$, is 
given analytically as $R_{L} = (1+\delta_{v})^{1/3}\bar{R}_{E}$. 
As can be seen the analytic and the numerical results are in excellent 
quantitative  agreements for all cases, proving the validity of our model.
It is worth emphasizing here that our model is completely analytical without 
having any fitting parameter derived from principles. It depends only on the 
initial cosmological parameters.

{\it Discussion.---}The ellipticity distribution of galaxy clusters has been 
used recently to constrain cosmology \cite{jin-sut02,lee06,ho-etal06}. 
For the case of clusters, however, the ellipticity distribution cannot be 
evaluated analytically from the initial conditions because the nonlinear 
clustering process tends to modify the cluster ellipticities significantly 
in the subsequent evolution. Although \cite{lee-etal05} attempted to derive 
an analytic fitting model for the cluster ellipticity distribution, their 
model was shown to be not in good quantitative agreements with the numerical 
results.

In contrast, we have for the first time shown here that the void ellipticity 
distribution can be evaluated fully analytically using physical principles and 
thus be a better indicator of the initial conditions of the universe since 
the void ellipticities are less vulnerable to nonlinear modifications.  
By applying our analytic model to real data that will be available from 
future large galaxy surveys , it may be possible to constrain the cosmological 
parameters precisely in an independently way. 

It is, however, worth nothing that the Hoyle-Vogeley algorithm works very 
well if a void is found as a spherical low-density region. Nevertheless, we 
expect that our numerical result would not change sensitively with respect 
to the void-finding algorithm, given the fact that the ellipticity of a void 
here is defined in terms of the spatial distribution of galaxies that make 
up the void but not by the boundary shape of a void.

The Millennium Run simulation used in this paper was carried out by the Virgo 
Supercomputing Consortium at the Computing Centre of the Max-Planck Society 
in Garching. We thank the referee and V. Springel for useful comment. 
This work was supported by the New Faculty Settlement fund of the Seoul 
National University.


\begin{thebibliography}{}
\bibitem{col-etal01}
Colless {\it et al.},  Mon. Not. R. Astron. Soc. {\bf 328}, 1039
(2001)
\bibitem{hoy-vog04}
F. Hoyle, and M. S. Vogeley, Astrophys. J. {\bf 607}, 751 (2004)
\bibitem{ick84}
V. Icke, Mon. Not. R. Astron. Soc. {\bf 206}, 1P (1984)
\bibitem{dub-etal93}
J. Dubinski, L. N. da Costa, D. S. Goldwirth, M. Lecar, and T. Piran, 
Astrophys. J. {\bf 410}, 458 (1993)
\bibitem{van-van93}
Van de Weygaert, R., and Van Kampen, E., Mon. Not. R. Astron. Soc.
{\bf 263}, 481 (1993)
\bibitem{she-van04}
R. K. Sheth, and R. van de Weygaert, Mon. Not. R. Astron. Soc. {\bf
350}, 517 (2004)
\bibitem{sha-etal04}
S. F. Shandarin, J. V. Sheth, and V. Sahni, Mon. Not. R.
Astron. Soc. {\bf 353}, 162 (2004)
\bibitem{sha-etal06}
S. F. Shandarin, H. A. Feldman, K. Heitmann, and S. Habib, Mon. Not. R.
Astron. Soc. {\bf 367}, 1629 (2006)
\bibitem{lee-par06}
J. Lee, and D. Park, Astrophys. J. {\bf 652}, 1 (2006)
\bibitem{zel70}
Ya. B. Zel'dovich, Astrophys. \& Astron. {\bf 5}, 84 (1970)
\bibitem{dor70}
A. G. Doroshkevich, Astrofizika, {\bf 3}, 175 (1970)
\bibitem{bar-etal86}
J. M. Bardeen, J. R. Bond, N. K. Kaiser, and  A. S. Szalay,  
Astrophys. J. {\bf 304}, 15 (1986)
\bibitem{lee-etal05} 
J. Lee, Y. Jing, and Y. Suto, Astrophys. J. {\bf 632}, 706 (2005)
\bibitem{spr-etal05}
V. Springel {\it et al.}, Nature, {\bf 435}, 629 (2005). 
The semianalytic galaxy catalogue is publicly available at 
http://www.mpa-garching.mpg.de/galform/agnpaper
\bibitem{hoy-vog02}
F. Hoyle, and M. S. Vogeley, Astrophys. J. {\bf 566}, 641 (2002)
\bibitem{jin-sut02}
Y. Jing, and Y. Suto, Astrophys. J. {\bf 574}, 538 (2002)
\bibitem{lee06}
J. Lee, Astrophys. J. {\bf 643}, 724 (2006)
\bibitem{ho-etal06}
Ho. Shirley, Bahcall, N. A., and P. Bode, Astrophys. J. {\bf 647}, 8 (2006)
\end{thebibliography}
\end{document}